\documentclass[apjl]{emulateapj}
\usepackage{apjfonts}

\shorttitle{Outflows in Post-starburst Galaxies}
\shortauthors{Tremonti et al.}

\begin{document}
\newcommand{\mg}{\ion{Mg}{2}~}
\newcommand{\kms}{km~s$^{-1}$}

\title{The Discovery of 1000 \kms{} Outflows in Massive Post-starburst Galaxies
at z=0.6\footnotemark[1]}

\author{Christy A. Tremonti\altaffilmark{2,3},
John Moustakas \altaffilmark{4},
Aleksandar M. Diamond-Stanic\altaffilmark{2},
}

\footnotetext[1]{Observations reported here were obtained at the MMT
Observatory, a joint facility of the University of Arizona and the
Smithsonian Institution.}
\altaffiltext{2}{Steward Observatory, 933 N.\ Cherry Ave., Tucson, AZ
  85721; tremonti@as.arizona.edu, adiamond@as.arizona.edu}
\altaffiltext{3}{Hubble Fellow}
\altaffiltext{4}{Center for Cosmology and Particle Physics, New York
  University, New York, NY 10003; jmoustakas@cosmo.nyu.edu}

\begin{abstract}

Numerical simulations suggest that active galactic nuclei (AGNs) play
an important role in the formation of early-type galaxies by expelling
gas and dust in powerful galactic winds and quenching star formation.
However, the existence of AGN feedback capable of halting galaxy-wide
star formation has yet to be observationally confirmed.  To
investigate this question, we have obtained spectra of 14
post-starburst galaxies at z$\sim$0.6 to search for evidence of
galactic winds.  In 10/14 galaxies we detect \mg
$\lambda\lambda2796,2803$ absorption lines which are blueshifted by
490 -- 2020~\kms with respect to the stars.  The median blueshift is
1140~\kms.  We hypothesize that the outflowing gas represents a fossil
galactic wind launched near the peak of the galaxy's activity, a few
100 Myr ago.  The velocities we measure are intermediate between those
of luminous starbursts and broad absorption line quasars, which
suggests that feedback from an AGN may have played a role in expelling
cool gas and shutting down star formation.

\end{abstract}


\section{Introduction}

There is mounting evidence linking quasar activity to merger-induced
star formation, but the precise timing and the physical relationship
between the two are not well understood \citep{Canalizo_et_al_2006}.
Numerical simulations suggest that mergers of gas-rich galaxies induce
radial gas inflows which fuel central star formation and black hole
accretion. Subsequently, feedback from the active galactic nucleus (AGN)
removes the gas and dust and quenches star formation and black hole
activity \citep[e.g.,][]{Di_Matteo_et_al_2005}.  Such models have
enjoyed great popularity due to their success in reproducing the
present-day properties of early-type galaxies, notably the
color--magnitude relation and the correlation between black hole mass
and bulge stellar velocity dispersion \citep[e.g.,][]{Granato_et_al_2004,
Springel_et_al_2005, Menci_et_al_2006}.  However, the existence of
AGN feedback capable of halting galaxy-wide star formation has yet to
be observationally confirmed.

AGN feedback is predicted to quench star formation by re-heating the
cold gas and expelling much of it in powerful galactic winds.
Galactic winds with velocities of 50 - 500 km~s$^{-1}$ are commonly
detected in starburst galaxies via the presence of gas absorption
lines that are blueshifted relative to stellar features
\citep[e.g.,][]{Heckman_et_al_2000}.  AGN-driven winds are expected to
produce similar observational signatures, but higher outflow
velocities \citep{Thacker_et_al_2006}.  The maximum feedback impulse
is predicted to occur during the bright quasar phase.  However, at
this stage the quasar outshines the host galaxy, and provides a more
ambiguous probe of the galaxy's interstellar medium (ISM).  (A
parsec-scale cloud near the quasar would be indistinguishable from a
kiloparsec-scale galactic wind.)  We have therefore elected to look
for remnants of AGN-driven galactic winds during the post-starburst
phase, a few 100 Myr after the peak of the star formation and AGN
activity.

Post-starburst galaxies are characterized by strong stellar Balmer
absorption from A-stars, but little nebular emission indicative of
on-going star formation.  Local post-starbursts (sometimes called
`E+A' or `K+A' galaxies) have the kinematic and morphological
signatures of pressure-supported spheroids, but frequently exhibit low
surface brightness tidal tails indicative of a recent major merger
\citep{Zabludoff_et_al_1996, Norton_et_al_2001, Yang_et_al_2004} and
signs of weak AGN activity \citep{Yan_et_al_2006, Yang_et_al_2006}.
Post-starbursts are therefore presumed to be late-stage mergers that
have passed through their quasar phase and are in transition to
becoming early-type galaxies.  As such they provide ideal testing
grounds for AGN-feedback models.

We have obtained spectra of 14 post-starbursts at $z\sim0.6$ in order
to search for evidence of galactic winds that may have played a role
in shutting down star formation.  We have selected galaxies at
intermediate redshift because this may be an important epoch for the
formation of early-type galaxies \citep[e.g.,][]{Faber_et_al_2005},
and because the rest-frame near-UV is accessible in the optical.
Coverage of the near-UV improves our ability to estimate the recent
star formation history of our galaxies and it enables us to measure
\mg $\lambda\lambda 2796,2803$ which is a sensitive probe of the ISM.
We describe our observations in \S2, the stellar populations of our
post-starbursts in \S3, and the outflow kinematics in \S4.  We discuss
evidence that the wind is powered by an AGN in \S5 and conclude in
\S6.

\section{Observations and Data Reduction}

Our sample was selected from the Sloan Digital Sky Survey (SDSS) Data
Release 4 \citep{Adelman_McCarthy_et_al_2006}.  The parent sample is
composed of $i < 20.5$ mag objects that were targeted for SDSS
spectroscopy as quasar candidates, but which were subsequently
classified as galaxies at $z$= 0.5 -- 1.  The typical signal-to-noise
(S/N) ratio of the SDSS spectra is rather poor
(S/N$\sim$2~pixel$^{-1}$), but sufficient for us to select a sample of
galaxies for follow-up.  We selected 159 objects with
post-starburst characteristics --- strong stellar Balmer absorption
and weak nebular emission (Tremonti et al., in prep.).

\begin{figure*}
\begin{center}
\leavevmode
\epsscale{0.92}
\plotone{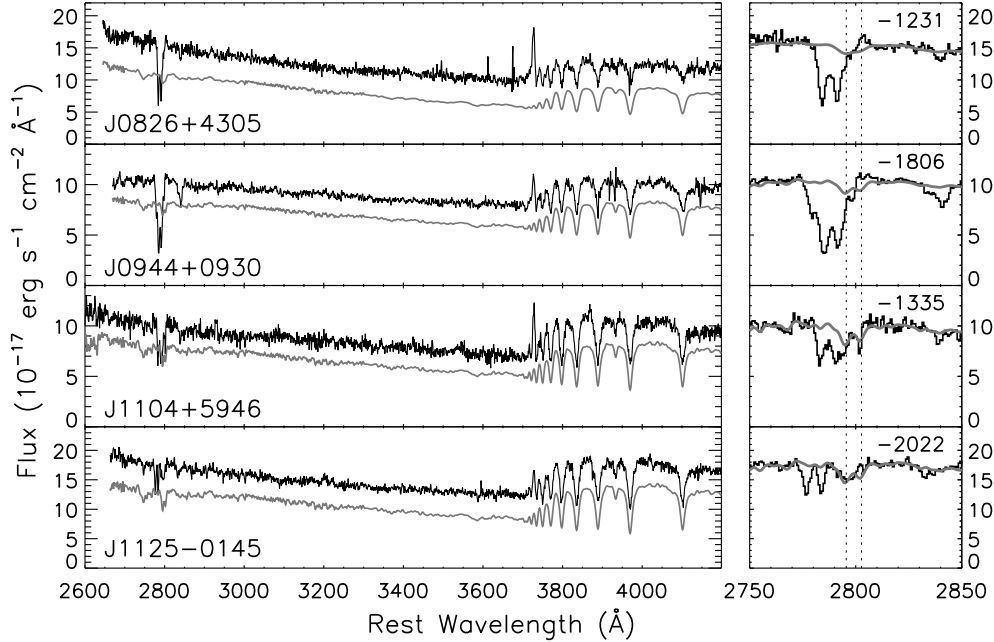}
\caption{Example spectra (black) and continuum model fits (gray).  In
the lefthand panel, the continuum models are offset for clarity. The
righthand panel highlights the region around the \mg doublet.  Dotted
lines mark the rest wavelength of \mg. The presence of blueshifted lines
indicates an outflow.  The velocity of the most blueshifted component
is given in \kms in the upper right corner.}
\end{center}
\end{figure*}

We obtained higher S/N spectra of 14 galaxies with the Blue Channel
Spectrograph on the 6.5-m MMT.  We used the $500$~line~mm$^{-1}$
grating blazed at $5600$~\AA\ which gave us spectral coverage from
$4050-7200$~\AA\ with a dispersion of $1.19$~\AA\ per pixel.  For our
$z=0.51 - 0.75$ galaxies, this yielded rest-frame coverage from 2700 -
4100~\AA.  Because most of our objects are unresolved in the SDSS
imaging we used a $1\arcsec$ slit, which yielded a FWHM resolution of
$\sim3.6$~\AA.  The spectra were reduced, extracted, and
spectrophotometrically calibrated using the ISPEC2D data reduction
package \citep{Moustakas_Kennicutt_2006}.  The MMT spectra show good
agreement with the SDSS data, but have S/N = 15-30
pixel$^{-1}$. Spectra of four representative galaxies are shown in
Figure~1.

\section{Stellar Population}  

We detect significant \mg $\lambda\lambda 2796,2803$ absorption in all
of our galaxies. \mg is one of the strongest interstellar resonance
absorption lines; however, it is also present in the atmospheres of
A-stars and later spectral types. Hence to accurately measure the ISM
absorption lines, we must carefully model the stellar continuum.  We
use the \citet[][hereafter BC03]{Bruzual_Charlot_2003} stellar
population synthesis models to create synthetic spectra for different
star formation histories.  We assume super-solar metallicity
(Z=2.5~Z$_{\sun}$) since Oxygen abundances several times solar are
measured in comparably luminous star forming galaxies at $z\sim0.6$
\citep{Lamareille_et_al_2006}.  We adopt a star formation history
designed to emulate a major merger between gas-rich disk galaxies.
Following the starburst, star formation decays exponentially with time
constants ranging from $\tau_{burst}$=25 -- 500 Myr.  We fit each of
our spectra with a grid of models spanning a range of ages,
$\tau_{burst}$, and reddening values and adopt the model with the
minimum $\chi^2$ as the best fit.  In several cases to achieve an
optimal fit it was necessary to add an additional power-law component
which may represent a featureless quasar continuum (see \S5).

A byproduct of our stellar continuum modeling is an estimate of the
stellar mass, the time since the peak star formation event
($t_{burst}$), and how quickly star formation ceased
($\tau_{burst}$).  These parameters and their uncertainties will be
discussed fully in Tremonti et al., in prep. Our modeling suggests
that the galaxies are massive ($0.7 - 4.8\times10^{11}$~M$_{\sun}$)
and have recently experienced a burst ($t_{burst} = 75 - 300$~Myr)
that faded rapidly ($\tau_{burst} =  25 - 100$~Myr).  The short
starburst timescales imply strong feedback from supernovae or an AGN.

\section{Gas Kinematics}  

We use our best-fit synthetic spectra to correct for the contribution
of stellar absorption to the \mg lines. At wavelengths less than
3300~\AA, the BC03 models use the \citet{Pickles_1998} stellar library
which has a spectral resolution of 10~\AA. This resolution is too low
to adequately model the \mg doublet in our data. We circumvent this
problem by patching the BC03 models in the 2600--3300 \AA\ range
using theoretical stellar spectra from the UVBLUE stellar library
\citep{Rodriguez-Merino_et_al_2005}. The rightmost panel in Figure~1
shows our continuum fits in the 2750--2850 \AA\ region. In many cases
the ISM lines are so strong or blueshifted that the stellar component
of \mg is unimportant. However, in a few galaxies stellar \mg is
dominant.  After correcting for the stellar light we find that 10 of
our 14 galaxies have measurable \mg absorption. The equivalent widths
(EWs) of interstellar \mg range from 0.8 -- 10.4~\AA\ (see Table~1).

After correcting for the stellar contribution to \mg, we fit the ISM
absorption lines following \citet{Rupke_et_al_2005a}.  In the
optically thin case the doublet ratio is 2:1, but in our data the
lines are moderately saturated.  At our spectral resolution
($\sim100$~\kms) this produces degeneracies between the optical depth
at line center, the covering factor, and the Doppler $b$
parameter. However, velocities can be measured robustly.  The lines
have a median Doppler width of $b=260$~\kms, although it is possible
that they include narrower unresolved components.  In six of the
galaxies we fit two absorption components.  Three galaxies display a
P-Cygni profile --- blueshifted absorption coupled with redshifted
emission which may originate on the back side of an expanding shell.
We model the emission with a Gaussian. In Table~1 we list the measured
absorption-line velocities.  We denote the velocity of the most
blueshifted component in each spectrum as v$_{max}$.  We compute the
average velocity, v$_{avg}$, weighting the components by their EWs.
The median values for the sample are v$_{avg} = -920$~\kms and
v$_{max} = -1140$~\kms.

We hypothesize that the blueshifted \mg lines originate in galactic
winds that were launched near the peak of the galaxies' starburst
activity a few 100~Myr ago.  An alternate interpretation is that the
absorbing gas is tidal debris associated with the merger. However, the
inner parts of tidal tails are expected to be bound and to fall back
within a few 100 Myr \citep{Hibbard_and_Mihos_1995}, whereas we detect
outflows.  In addition, gaseous tidal tails are confined to relatively
thin streams with small global covering factors; therefore it seems
unlikely that we would detect tidal gas in absorption in 70\% of our
sources.  Hence we conclude that the blueshifted \mg lines originate in
fossil galactic winds.

\section{Discussion}

Our $z\sim0.6$ post-starburst galaxies rank among the most luminous
and massive galaxies in the universe (M$_B=$ -22.5 -- -23.7~mag, M$_{*} =
0.7 - 4.8 \times 10^{11}$~M$_{\sun}$; Table~1) and they offer a rare
window on the formation of today's massive early-type galaxies.  The
detection of interstellar \mg in 10 of our 14 galaxies enables us to
probe the properties of the cold ISM.  We find evidence for strong
outflows in all 10 systems, with velocities in the range v$_{max}=500$
-- 2000~\kms. The median $v_{max}$ of the sample is 1140 \kms, which
exceeds the 400 -- 600 \kms velocities typical of luminous starburst
galaxies \citep{Heckman_et_al_2000}.  We put these outflows in context
in Figure~2 where we plot absolute $B$-band magnitude versus ISM
outflow velocity for a variety of systems.  We include local
starbursts \citep{Schwartz_Martin_2004, Schwartz_et_al_2006}, Luminous
and Ultra-Luminous Infrared Galaxies
\citep[LIRGs/ULIRGs;][]{Rupke_et_al_2005b}, and $z\sim3$ Lyman Break
Galaxies \citep[LBGs;][]{Pettini_et_al_2001}.  We augment this sample
with starburst/AGN composite ULIRGs from \citet{Rupke_et_al_2005c} and
a sample of Low-ionization Broad Absorption Line quasars (LoBALs) from
the SDSS \citep{Trump_et_al_2006}.  LoBALs are characterized by broad
\mg absorption troughs.  They are more common in infrared-selected
than optically-selected quasar samples \citep{Boroson_Meyers_1992}, which
has led to the suggestion that LoBALs are quasars in the process of 
removing their natal cocoons of gas and dust.

Figure~2 shows a striking trend for more luminous galaxies to have
higher outflow velocities.  Similar trends have been noted previously.
In starbursts, \citet{Rupke_et_al_2005b} and \citet{Martin_2005} found
strong correlations between outflow velocity and galaxy mass and star
formation rate.  Our post-starburst galaxies have extraordinarily high
outflow velocities when compared to their natural analogs,
starburst-powered ULIRGs and LBGs. Their outflow velocities are
comparable to some of the AGN composite ULIRGs, and at the lower end
of the range observed for LoBAL quasars.  This result implies that our
post-starburst galaxies may harbor both fading starbursts and fading
quasars.  Evidence for the presence of an AGN can also be found in the
spectra.  In massive metal-rich galaxies the narrow emission line
[\ion{O}{3}] $\lambda5007$ is relatively uncontaminated by star
formation and a good tracer of the AGN's bolometric luminosity
\citep{Heckman_et_al_2005}.  We are able to detect [\ion{O}{3}] lines
in four of our galaxies using the SDSS spectra.  The galaxies have
EW$_{\mathrm{[OIII]}} = 6 - 9$~\AA\ and L$_{\mathrm{[OIII]}} = 0.4 - 4 \times
10^8$~L$_{\sun}$, placing them in the regime of powerful AGN.  One of
the four also shows [\ion{Ne}{5}]~$\lambda3426$ emission, which is an
unequivocal signpost of AGN activity.

\begin{figure}
\epsscale{1.0}
\plotone{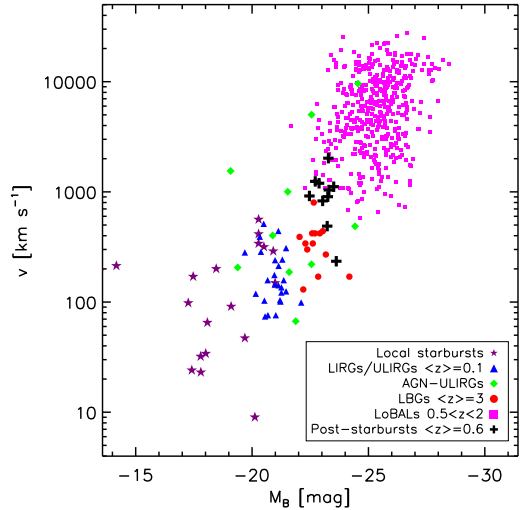}
\caption{Outflow velocity versus absolute $B$-band magnitude. Our 
post-starburst galaxies (black plus signs) have velocities intermediate 
between luminous starbursts and LoBAL quasars. References for the various 
samples are given in the text.}
\end{figure}

We achieve better continuum fits to six of our galaxies (three of
which have [\ion{O}{3}] emission) when we add a featureless power-law
component ($F_{\lambda} \propto \lambda^{\alpha_{\lambda}}$) with a
spectral slope of $\alpha_{\lambda}=-1.6$, which is typical of quasars
\citep{Vanden_Berk_et_al_2001}.  Without the power-law, high
present-day star formation rates are implied, which is at odds with
the lack of strong Balmer emission.  The very blue UV continua of
these galaxies rules out high dust attenuation as a means of quenching
the nebular lines. In our best-fit models the power-law supplies
40--70\% of the flux at 3000~\AA.  Curiously, despite the moderate
dilution of the AGN continuum by the galaxy, few spectral features
characteristic of Type 1 quasars are detected.  We are able to rule
out strong broad H$\beta$ emission lines on the basis of the SDSS
spectra.  Broad \mg (FWHM$\sim$8000~\kms) is present at a low level in
SDSS~J214000.49+120914.6, but absent in the other spectra.  The
physical reason for the lack of broad lines in our AGN-post-starburst
composites is unclear.

Mechanical energy from AGN radio jets has been suggested as a power
source for large-scale outflows \citep[e.g.,][]{Nesvadba_et_al_2006}.
For 13/14 galaxies, radio data are available from the Faint Images of
the Radio Sky at Twenty cm survey \citep[FIRST;][]{Becker_et_al_1995}.
Two galaxies are detected ($F_{\mathrm{1.4 GHz}} = 12$, 20~mJy) down
to a limit of $\sim1$~mJy.  These two sources show no optical signs of
AGN activity, but have sufficient radio power to be classed as
radio-loud AGN \citep{Kellermann_et_al_1989}.  Neither galaxy hosts an
outflow, thus, preliminary evidence disfavors radio jets as the
driving mechanism.

It is interesting to consider whether the outflows we observe could
have been the direct cause of the abrupt drop in the star formation
rate of our galaxies a few 100 Myr ago.  This seems plausible if the
winds entrained a large fraction of the cold ISM.  We can crudely
estimate the mass in the outflow using the \mg column density that we
derive.  Because the error bars on some individual measurements
are large, we use the median value, N(Mg$^{+}$) = $8.1\times
10^{14}$~cm$^{-2}$, in our calculations.  The presence of a weak
\ion{Mg}{1}~$\lambda2852$ line in some of the spectra implies an
ionization correction of $\sim$3\%.  We translate our Mg column into a
total gas column by accounting for depletion onto dust grains
\citep[X=-1.4;][]{Savage_Sembach_1996} and assuming a Mg/H ratio 2.5
times the solar value. In this way we infer N(H)=$2 \times
10^{20}$~cm$^{-2}$, which is consistent with the median value found
for high-$z$ ULIRGs \citep{Rupke_et_al_2005b}.

The total mass in the outflow depends strongly on how far away the
absorbing gas is from the galaxy.  Following
\citet{Rupke_et_al_2005b}, we assume that the wind is a shell-like
structure that covers 80\% of the optically luminous galaxy.  To
calculate the outer radius of the shell, $r_{out}$, we adopt a simple
picture where the wind is launched at the peak of the star formation
and AGN activity and moves at constant velocity.  Using the median
values of the burst age ($t\simeq100$~Myr) and outflow velocity
(v$\simeq1000$~\kms), we find $r_{out} = v t = 100$~kpc.  The shell's
thickness depends on the duration of the outflow.  We consider two
limiting cases: a thick shell with an inner radius $r_{in}=1$~kpc, and
a thin shell with $r_{in}=99$~kpc.  Rupke's equation (4) yields masses
of $M_{wind} = 10^{9}$ and $10^{11}$~M$_{\sun}$ for the thick and thin
cases respectively, implying that between 1 and 50\% of the galaxies'
baryons are in the outflow.  The wind mass estimated for the thin
shell is probably too large: simulations of starburst and AGN feedback
require highly efficient and energetic winds to unbind more than 25\%
of the galaxy's initial gas mass (Cox et al., in prep).
 
Another possibility we must consider is that the \mg absorber is local
to the AGN and does not extend to kiloparsec scales.  In this case the
wind is unlikely to have played a significant role in regulating star
formation.  For six of our galaxies this possibility cannot be ruled
out because our estimate of the \mg covering factor is less than or
equal to the amount of continuum light contributed by the AGN.  We
obtained data with $5\times$ higher spectral resolution for SDSS
J082638.41+430529.5 in order to obtain a better measurement of the
covering factor. In the high resolution spectrum, the \mg absorption
is near-black at line center, implying that the absorber covers both
the AGN and the stars.  We also detect strong \mg absorption in three
galaxies with no contribution to the continuum from an AGN.  Hence, in
4/10 galaxies we can confirm that the winds are galaxy-scale features
indicative of energetically significant feedback events.  Constraints
on the remaining galaxies await higher resolution spectra.

\section{Conclusions}

To test currently popular models of early-type galaxy evolution that
incorporate feedback from AGN, we have looked for the presence of
galactic winds in a sample of massive post-starburst galaxies at
$z=0.5 - 0.75$.  We detect interstellar \mg which is blueshifted by
500 - 2000 \kms in 10/14 galaxies. These outflow velocities are
intermediate between those of luminous starbursts and LoBAL quasars,
which suggests that feedback from an AGN may have played a role in
powering the outflow. In 4/10 galaxies we can confirm that the
outflows are energetically significant galaxy-wide events, and not
phenomena local to the AGN.  We estimate that the outflows reach
distances of $\sim$100~kpc and contain upwards of $10^9$~M$_{\sun}$ of gas.
We conclude that AGN are likely to have played a major role in causing
the abrupt truncation of star formation in these massive galaxies.

\acknowledgements

We thank Tim Heckman for helpful discussions and Kevin Luhman for
contributing telescope time.  We are grateful to the Aspen Center for
Physics for hospitality while part of this work was completed.
Support for C. A. T. was provided by NASA through Hubble Fellowship
grants HST-HF-01192.01 awarded by the Space Telescope Science
Institute, which is operated by the Association of Universities for
Research in Astronomy, Inc., for NASA, under contract NAS5-26555.

Funding for the Sloan Digital Sky Survey (SDSS) has been provided by
the Alfred P. Sloan Foundation, the Participating Institutions, the
National Aeronautics and Space Administration, the National Science
Foundation, the U.S. Department of Energy, the Japanese
Monbukagakusho, and the Max Planck Society. The SDSS Web site is
http://www.sdss.org/.




\begin{deluxetable}{cccccrr}
\tablehead{
\colhead{SDSS Galaxy Name} &
\colhead{$z$} &
\colhead{M$_{B}$} &
\colhead{log M$_{*}$} &
\colhead{\mg EW} &
\colhead{v$_{avg}$} &
\colhead{v$_{max}$}
\\
\colhead{ } &
\colhead{ } &
\colhead{(mag)} &
\colhead{(M$_{\sun}$)} &
\colhead{(\AA)} &
\colhead{(\kms)} &
\colhead{(\kms)}
}
\startdata
J081150.09+471615.2  &   0.515  &  -22.5  &   11.0  &    2.5  &  -918 $\pm$ 27
  &  -918 $\pm$ 27 \\
J082638.41+430529.5  &   0.603  &  -23.3  &   10.9  &    4.8  & -1040 $\pm$ 42
  & -1232 $\pm$ 09 \\
J082733.88+295451.3  &   0.681  &  -23.1  &   11.3  &\nodata  &\nodata  &\nodata \\
J094417.85+093019.4  &   0.514  &  -22.7  &   10.8  &    7.9  & -1245 $\pm$ 10
  & -1807 $\pm$ 13 \\
J103906.97+453754.1  &   0.634  &  -23.3  &   11.1  &    4.1  &  -904 $\pm$ 31
  & -1342 $\pm$ 40 \\
J110437.46+594639.6  &   0.573  &  -22.9  &   10.9  &    3.4  & -1197 $\pm$ 35
  & -1335 $\pm$ 22 \\
J112518.90-014532.5  &   0.519  &  -23.3  &   11.1  &    1.9  & -2022 $\pm$ 10
  & -2022 $\pm$ 10 \\
J114257.23+603711.2  &   0.568  &  -23.6  &   11.5  &\nodata  &\nodata  &\nodata \\
J124807.16+060111.8  &   0.632  &  -23.2  &   11.2  &    2.9  &  -489 $\pm$ 18
  &  -489 $\pm$ 18 \\
J150636.30+540220.9  &   0.608  &  -23.5  &   10.9  &    2.5  & -1114 $\pm$ 66
  & -1135 $\pm$ 78 \\
J160413.25+393931.4  &   0.564  &  -23.6  &   11.7  &\nodata  &\nodata  &\nodata \\
J163541.72+470924.5  &   0.699  &  -23.7  &   11.5  &\nodata  &\nodata  &\nodata \\
J171300.39+281708.2  &   0.577  &  -23.0  &   11.2  &    0.8  &  -828 $\pm$ 35
  &  -828 $\pm$ 35 \\
J214000.49+120914.6  &   0.751  &  -23.6  &   11.2  &   10.4  &  -234 $\pm$ 40
  &  -573 $\pm$ 48 \\
\enddata

\tablecomments{We assume $\Omega_{M} = 0.3$, $\Omega_{\Lambda} = 0.7$,
and H$_0$ = 70 km~s$^{-1}$ Mpc$^{-1}$.  M$_B$ is $k-$corrected 
to $z=0$ and on the Vega system.  The \mg EWs are for the interstellar
component and are measured in the rest frame. The velocities 
v$_{max}$ and v$_{avg}$ are defined in \S4.}

\end{deluxetable}


\begin{thebibliography}

\bibitem[Adelman-McCarthy et al.(2006)]{Adelman_McCarthy_et_al_2006} 
Adelman-McCarthy, J.~K., et al.\ 2006, \apjs, 162, 38 

\bibitem[Becker et al.(1995)]{Becker_et_al_1995} 
Becker, R.~H., White, R.~L., \& Helfand, D.~J.\ 1995, \apj, 450, 559 

\bibitem[Boroson \& Meyers(1992)]{Boroson_Meyers_1992} 
Boroson, T.~A., \& Meyers, K.~A.\ 1992, \apj, 397, 442 

\bibitem[Bruzual \& Charlot(2003)]{Bruzual_Charlot_2003} 
Bruzual, G., \& Charlot, S.\ 2003, \mnras, 344, 1000 

\bibitem[Canalizo et al.(2006)]{Canalizo_et_al_2006} 
Canalizo, G., Stockton, A., Brotherton, M.~S., \& Lacy, M.\ 2006, 
New Astronomy Review, 50, 650 

\bibitem[Di Matteo et al.(2005)]{Di_Matteo_et_al_2005} 
Di Matteo, T., Springel, V., \& Hernquist, L.\ 2005, \nat, 433, 604 

\bibitem[Faber et al.(2005)]{Faber_et_al_2005} 
Faber, S.~M., et al.\ 2005, ArXiv Astrophysics e-prints, 
arXiv:astro-ph/0506044 

\bibitem[Granato et al.(2004)]{Granato_et_al_2004} 
Granato, G.~L., De Zotti, G., Silva, L., Bressan, A., \& Danese, L.\
2004, \apj, 600, 580

\bibitem[Heckman et al.(2000)]{Heckman_et_al_2000} 
Heckman, T.~M., Lehnert, M.~D., Strickland, D.~K., \& Armus, L.\ 2000, 
\apjs, 129, 493 

\bibitem[Heckman et al.(2005)]{Heckman_et_al_2005} 
Heckman, T.~M., Ptak, A., Hornschemeier, A., \& Kauffmann, G.\ 2005, 
\apj, 634, 161 

\bibitem[Hibbard \& Mihos(1995)]{Hibbard_and_Mihos_1995}
Hibbard, J.~E.\ \& Mihos, J.~C.\ 1995, \aj, 110, 140

\bibitem[Kellermann et al.(1989)]{Kellermann_et_al_1989} 
Kellermann, K.~I., Sramek, R., Schmidt, M., Shaffer, D.~B., \& Green,
R.\ 1989, \aj, 98, 1195

\bibitem[Lamareille et al.(2006)]{Lamareille_et_al_2006} Lamareille, F., 
Contini, T., Brinchmann, J., Le Borgne, J.-F., Charlot, S., \& Richard, J.\ 
2006, \aap, 448, 907 

\bibitem[Martin(2005)]{Martin_2005} 
Martin, C.~L.\ 2005, \apj, 621, 227 

\bibitem[Menci et al.(2006)]{Menci_et_al_2006} 
Menci, N., Fontana, A., Giallongo, E., Grazian, A., \& Salimbeni, 
S.\ 2006, \apj, 647, 753 

\bibitem[Moustakas \& Kennicutt(2006)]{Moustakas_Kennicutt_2006} 
Moustakas, J., \& Kennicutt, R.~C., Jr.\ 2006, \apjs, 164, 81

\bibitem[Nesvadba et al.(2006)]{Nesvadba_et_al_2006} 
Nesvadba, N.~P.~H., Lehnert, M.~D., Eisenhauer, F., Gilbert, A., 
Tecza, M., \& Abuter, R.\ 2006, \apj, 650, 693 

\bibitem[Norton et al.(2001)]{Norton_et_al_2001} 
Norton, S.~A., Gebhardt, K., Zabludoff, A.~I., \& Zaritsky, D.\ 2001, 
\apj, 557, 150 

\bibitem[Pettini et al.(2001)]{Pettini_et_al_2001} 
Pettini, M., Shapley, A.~E., Steidel, C.~C., Cuby, J.-G., Dickinson,
M., Moorwood, A.~F.~M., Adelberger, K.~L., \& Giavalisco, M.\ 2001, 
\apj, 554, 981 

\bibitem[Pickles(1998)]{Pickles_1998} 
Pickles, A.~J.\ 1998, \pasp, 110, 863 

\bibitem[Rodr{\'{\i}}guez-Merino et al.(2005)]{Rodriguez-Merino_et_al_2005} 
Rodr{\'{\i}}guez-Merino, L.~H., Chavez, M., Bertone, E., \& Buzzoni, A.\ 
2005, \apj, 626, 411 

\bibitem[Rupke et al.(2005a)]{Rupke_et_al_2005a} 
Rupke, D.~S., Veilleux, S., \& Sanders, D.~B.\ 2005, \apjs, 160, 87 

\bibitem[Rupke et al.(2005b)]{Rupke_et_al_2005b} 
Rupke, D.~S., Veilleux, S., \& Sanders, D.~B.\ 2005, \apjs, 160, 115 

\bibitem[Rupke et al.(2005c)]{Rupke_et_al_2005c} 
Rupke, D.~S., Veilleux, S., \& Sanders, D.~B.\ 2005, \apj, 632, 751 

\bibitem[Savage \& Sembach(1996)]{Savage_Sembach_1996} 
Savage, B.~D., \& Sembach, K.~R.\ 1996, \araa, 34, 279 

\bibitem[Schwartz \& Martin(2004)]{Schwartz_Martin_2004} 
Schwartz, C.~M., \& Martin, C.~L.\ 2004, \apj, 610, 201 

\bibitem[Schwartz et al.(2006)]{Schwartz_et_al_2006} 
Schwartz, C.~M., Martin, C.~L., Chandar, R., Leitherer, C., 
Heckman, T.~M., \& Oey, M.~S.\ 2006, \apj, 646, 858 

\bibitem[Springel et al.(2005)]{Springel_et_al_2005} 
Springel, V., Di Matteo, T., \& Hernquist, L.\ 2005, \apjl, 620, L79 

\bibitem[Thacker et al.(2006)]{Thacker_et_al_2006} 
Thacker, R.~J., Scannapieco, E., \& Couchman, H.~M.~P.\ 2006, 
\apj, 653, 86 


\bibitem[Trump et al.(2006)]{Trump_et_al_2006} 
Trump, J.~R., et al.\ 2006, \apjs, 165, 1 

\bibitem[Vanden Berk et al.(2001)] {Vanden_Berk_et_al_2001} 
Vanden Berk, D.~E., et al.\ 2001, \aj, 122, 549 

\bibitem[Yan et al.(2006)]{Yan_et_al_2006} 
Yan, R., Newman, J.~A., Faber, S.~M., Konidaris, N., Koo, D., \&
Davis, M.\ 2006, \apj, 648, 281

\bibitem[Yang et al.(2004)]{Yang_et_al_2004} 
Yang, Y., Zabludoff, A.~I., Zaritsky, D., Lauer, T.~R., \& Mihos, 
J.~C.\ 2004, \apj, 607, 258 

\bibitem[Yang et al.(2006)]{Yang_et_al_2006} 
Yang, Y., Tremonti, C.~A., Zabludoff, A.~I., \& Zaritsky, D.\ 2006, 
\apjl, 646, L33 

\bibitem[Zabludoff et al.(1996)]{Zabludoff_et_al_1996}
Zabludoff, A.~I., Zaritsky, D., Lin, H., Tucker, D., Hashimoto, Y., 
Shectman, S.~A., Oemler, A., \& Kirshner, R.~P. 1996, \apj, 466, 104 

\end{thebibliography}
\end{document}